\documentclass[letter, noresetcount,conference, compsocconf]{llncs}
\usepackage{times}
\usepackage{epsfig}
\usepackage{amsmath}
\usepackage{amssymb}
\usepackage{verbatim}
\usepackage[noend,resetcount,boxed]{algorithm2e}
\usepackage{setspace}
\usepackage[active]{srcltx}
\usepackage{paralist}

\newcommand{\spara}[1]{\smallskip\noindent{\bf{#1}}}

\newcommand{\la}{\leftarrow}

\newcommand{\POabcast}{{\em POabcast}}

\newcommand{\tran}[1]{\stackrel{#1}{\rightarrow}}
\newcommand{\Tran}[1]{\stackrel{#1}{\Rightarrow}}

\newcommand{\barrier}{\tau}
\newcommand{\newepoch}{NEW-EPOCH}
\newcommand{\msgvalue}{VAL}

\newtheorem{thm}{\bf Theorem}
\newtheorem{defn}[thm]{\bf Definition}

\usepackage[hide]{chato-notes}

\usepackage{fullpage}

\begin{document}

\SetKw{Event}{upon}
\SetKwBlock{Do}{}{}
\SetKw{Thread}{thread}
\SetKw{Proc}{procedure}
\SetKw{Funct}{function}
\SetKwIF{If}{ElseIf}{Else}{if}{then}{else if}{else}{endif}

\newenvironment{itemize*}%
  {\begin{itemize}%
    \setlength{\itemsep}{0.2mm}%
    \setlength{\parskip}{0.2mm}}%
  {\end{itemize}}

\title{On Barriers and the Gap between Active and Passive Replication\\ (Full Version)\thanks{A conference version of this work (without appendices) appears in the proceedings of the 27th International Symposium on Distributed Computing (DISC) 2013 conference.}}

\author{
Flavio P. Junqueira\inst{1}
\and 
Marco Serafini\inst{2}
}
\institute{
Microsoft Research, Cambridge, UK\\ \email{fpj@apache.org} \and
Yahoo! Research, Barcelona, Spain\\ \email{serafini@yahoo-inc.com}
}
\date{}



\maketitle

\begin{abstract}

Active replication is commonly built on top of the atomic broadcast primitive. 
Passive replication, which has been recently used in the popular ZooKeeper coordination system, can be naturally built on top of the primary-order atomic broadcast primitive. 
Passive replication differs from active replication in that it requires processes to cross a {\em barrier} before they become primaries and start broadcasting messages. 
In this paper, we propose a barrier function $\barrier$ that explains and encapsulates the differences between existing primary-order atomic broadcast algorithms, namely semi-passive replication and Zookeeper atomic broadcast (Zab), as well as the differences between Paxos and Zab.
We also show that implementing primary-order atomic broadcast on top of a generic consensus primitive and $\barrier$ inherently results in higher time complexity than atomic broadcast, as witnessed by existing algorithms. 
We overcome this problem by presenting an alternative, primary-order atomic broadcast implementation that builds on top of a generic consensus primitive and uses consensus itself to form a barrier.
This algorithm is modular and matches the time complexity of existing $\barrier$-based algorithms.

\end{abstract}

\section{Introduction}
Passive replication is a popular approach to achieve fault tolerance in practical systems~\cite{primary-backup}.
Systems like ZooKeeper~\cite{zookeeper} or Megastore~\cite{megastore} use primary replicas to produce state updates or state mutations.
Passive replication uses two types of replicas: primaries and backups.
A primary replica executes client operations, without assuming that the execution is deterministic, and produces state updates.
Backups apply state updates in the order generated by the primary.
With active replication, by contrast, all replicas execute all client operations, assuming that the execution is deterministic.
Replicas execute a sequence of consensus instances on client operations to agree on a single execution sequence using atomic broadcast (abcast).
Passive replication has a few advantages such as simplifying the design of replicated systems with non-deterministic operations, {\em e.g.}, those depending on timeouts or interrupts.

It has been observed by Junqueira {\em et al.}~\cite{zab} and Birman {\em et al.}~\cite{vs-paxos} that using atomic broadcast for passive, instead of active, replication requires taking care of specific constraints.
State updates must be applied in the exact sequence in which they have been generated: if a primary is in state $A$ and executes an operation making it transition to state update $B$, the resulting state update $\delta_{AB}$ must be applied to state $A$.
Applying it to a different state $C \neq A$ is not safe because it might lead to an incorrect state, which is inconsistent with the history observed by the primary and potentially the clients.
Because a state update is the difference between a new state and the previous, there is a causal dependency between state updates.
Unfortunately, passive replication algorithms on top of atomic broadcast (\emph{abcast}) do not necessarily preserve this dependency: if multiple primaries are concurrently present in the system, they may generate conflicting state updates that followers end up applying in the wrong order.
Primary-order atomic broadcast (\POabcast) algorithms, like Zab~\cite{zab}, have additional safety properties that solve this problem.
In particular, it implements a {\em barrier}, the {\em isPrimary}  predicate, which must be crossed by processes that want to broadcast messages.

Interestingly, the only existing passive replication algorithm using consensus as a communication primitive, the semi-passive replication algorithm of Defago {\em et al.}~\cite{defago2004semi}, has linear time complexity in the number of concurrently submitted requests.
Recent algorithms for passive replication have constant complexity but they directly implement \POabcast\ without building on top of consensus~\cite{vs-paxos,zab}.

During our work on the ZooKeeper coordination system~\cite{zookeeper} we have realized that it is still not clear how these algorithms relate, and whether this trade-off between modularity and time complexity is inherent.
This paper shows that existing implementations of passive replication can be seen as instances of the same unified consensus-based \POabcast\ algorithm, which is basically an atomic broadcast algorithm with a barrier predicate implemented through a {\em barrier function} $\tau$ we define in this work.
The $\tau$ function outputs the identifier of the consensus instance a leader process must decide on before becoming a primary.



Existing algorithms constitute alternative implementations of $\barrier$;
the discriminant is whether they consider the underlying consensus algorithm as a black-box whose internal state cannot be observed.
Our $\barrier$-based algorithm exposes an inherent trade off.
We show that if one implements $\barrier$ while considering the consensus implementation as a black box, it is {\em necessary} to execute consensus instances sequentially, resulting in higher time complexity.
This algorithm corresponds to semi-passive replication.

If the $\barrier$ implementation can observe the internal state of the consensus primitive, we can avoid the impossibility and execute parallel instances. 
For example, Zab is similar to the instance of our unified algorithm that uses Paxos as the underlying consensus algorithm and implements the barrier by reading the internal state of the Paxos protocol.
We experimentally evaluate that using parallel instances almost doubles the maximum throughput of passive replication in stable periods, even considering optimizations such as batching.
Abstracting away these two alternatives and their inherent limitations regarding time complexity and modularity is one of the main observations of this paper.

Finally, we devise a {\em $\barrier$-free} \POabcast\ algorithm that makes this trade off unnecessary, since it enables running parallel consensus instances using an unmodified consensus primitive as a black box.
Unlike barrier-based algorithms, a process becomes a primary by proposing a special value in the next available consensus instances;
this value marks the end of the sequence of accepted messages from old primaries.
Table~\ref{tab:comp} compares the different PO abcast algorithms we discuss in our paper.

Our barrier-free algorithm shows that both active and passive replication can be implemented on top of a black-box consensus primitive with small and well understood changes and without compromising performance.

\spara{Differences with the DISC'13 version.} This version includes gap handling in Algorithm~\ref{alg:PB-barrier}.
In addition, Section~\ref{sec:white-box} gives more details on how Paxos handles gaps.

\begin{table}[t]
\begin{center}

\caption{ Time complexity of \POabcast\ algorithms presented in this paper - see Sect.~\ref{sec:barrier-compl} and~\ref{sec:barrier-free-compl} for detail. We consider the use of Paxos as the underlying consensus algorithm since it has optimal latency~\cite{lower}. However, only the third solution {\em requires} the use of Paxos; the other algorithms can use any implementation of consensus. For the latency analysis only, we assume that message delays are equal to $\Delta$. The {\em Stable periods} column reports the time, in a passive replication system, between the receipt of a client request and its delivery by a single broadcasting primary/leader ($c$ is the number of clients). The {\em Leader change} column reports idle time after a new single leader is elected by $\Omega$ and before it can broadcast new  messages.}
\label{tab:comp}

\begin{tabular}{|c|c|c|c|}
\hline 
  & Stable periods & Leader change\\ 
\hline 
Atomic broadcast \cite{paxos} & $2 \Delta$ & $2 \Delta$ \\ 
\hline 
$\barrier$-based \POabcast\ (Sect.~\ref{sec:black-box})  & $2\Delta \cdot c$ & $4 \Delta$ \\ 
\hline 
$\barrier$-based \POabcast\ with white-box Paxos (Sect.~\ref{sec:white-box}) & $2 \Delta$ & $4 \Delta$\\ 
\hline 
$\barrier$-free \POabcast\ (Sect.~\ref{sec:simulation}) & $2 \Delta$ & $4\Delta$ \\ 
\hline 
\end{tabular} 

\end{center}
\end{table}

\section{Related Work}
\label{sec:relwork}

Traditional work on passive replication and the primary-backup approach assumes synchronous links~\cite{primary-backup}.
Group communication has been used to support primary-backup systems; it assumes a $\Diamond P$ failure detector for liveness~\cite{memb-surv}.
Both atomic broadcast and \POabcast\ can be implemented in a weaker system model, {\em i.e.}, an asynchronous system equipped with an $\Omega$ leader oracle~\cite{weakest-CHT}.
For example, our algorithms do not need to agree on a new view every time a non-primary process crashes.

Some papers have addressed the problem of reconfiguration: dynamically changing the set of processes participating to the state machine replication group.
Vertical Paxos supports reconfiguration by using an external master, which can be a replicated state machine~\cite{vertical-paxos}.
This supports {\em primary-backup} systems, defined as replicated systems where write quorums consist of all processes and each single process is a read quorum.
Vertical Paxos does not address the issues of passive replication and considers systems where commands, not state updates, are agreed upon by replicas.
Virtually Synchronous Paxos (VS Paxos) aims at combining virtual synchrony and Paxos for reconfiguration~\cite{vs-paxos}.
Our work assumes a fixed set of processes and does not consider the problem of reconfiguring the set of processes participating to consensus.
Shraer {\em et al.} have recently shown that reconfiguration can be implemented {\em on top} of a \POabcast\ construction as the ones we present in this paper, making it an orthogonal topic~\cite{zk-reconf}.

While there has been a large body of work on group communication, only few algorithms implement passive replication in asynchronous systems with $\Omega$ failure detectors: semi-passive replication~\cite{defago2004semi}, Zab~\cite{zab} and Virtually synchronous Paxos~\cite{vs-paxos}.
We relate these algorithms with our barrier-based algorithms in Sect.~\ref{sec:existing}.

Pronto is an algorithm for database replication that shares several design choices with our $\barrier$-free algorithm and has the same time complexity in stable periods~\cite{pronto}.
Both algorithms elect a primary using an unreliable failure detector and have a similar notion of epochs, which are associated to a single primary.
Epoch changes are determined using an agreement protocol, and values from old epochs that are agreed upon after a new epoch has been agreed upon are ignored.
Pronto, however, is an active replication protocol: all replicas execute transactions, and non-determinism is handled by agreeing on a per-transaction log of non-deterministic choices that are application specific.
Our work focuses on passive replication algorithms, their difference with active replication protocols, and on the notion of barriers in their implementation.

\section{System Model and Primitives}
\label{sec:model}

Throughout the paper, we consider an asynchronous system composed of a set $\Pi = \{ p_1, \ldots, p_n \}$ of processes that can fail by crashing.
They implement a passive replication algorithm, executing requests obtained by an unbounded number of client processes, which can also fail by crashing.
{\em Correct} processes are those that never crash.
Processes are equipped with an $\Omega$ failure detector oracle.

\begin{defn}[Leader election oracle]~A leader election oracle $\Omega$ operating on a set of processes $\Pi$ outputs the identifier of some process $p \in \Pi$.
Instances of the oracle running on different processes can return different outputs. 
Eventually, all instances of correct processes permanently output the same correct process.
\end{defn}

Our algorithms build on top of (uniform) consensus, which has the following properties.

\begin{defn}[Consensus]~A consensus primitive consists of two operations: {\em propose}($v$) and {\em decide}($v$) of a value $v$. 
It satisfies the following properties:
\begin{itemize}
\item[\textbf{\em Termination.}] If some correct process proposes a value, every correct process eventually decides some value.
\item[\textbf{\em Validity.}] If a processes decides a value, this value was proposed by some process.
\item[\textbf{\em Integrity.}] Every correct process decides at most one value.
\item[\textbf{\em Agreement.}] No two processes decide differently.
\end{itemize}
\label{def:consensus}
\end{defn}

Since our algorithms use multiple instances of consensus, {\em propose} and {\em decide} have an additional parameter denoting the identifier of the consensus instance.

\begin{algorithm}[t]

\begin{small}

\textbf{initially} \Do{
	replied($c$, $t$) returns false for all inputs\; 
	$initialized \la false$\;
}

\Event isPrimary() \Do{
	$\Theta \la \Sigma$\;
	$initialized \la true$\;
}

\Event $\neg$isPrimary() \Do{
	$initialized \la false$\;
}

\Event receive $\langle c, t, o \rangle$ $\wedge$ $\neg$ replied($c$, $t$) $\wedge$ isPrimary() $\wedge$ $initialized$ \label{ln:alt1}\Do{
	$\Theta \tran{o} \langle r, \delta \rangle$\;
	{\em apply}$(\delta, \Theta)$\;
	POabcast ($\langle \delta, r, c, t\rangle$)\;
}

\Event receive operation $\langle c, t, o \rangle$ $\wedge$ replied($c$, $t$) \label{ln:alt1}\Do{
    	send stored $\langle c$, $t$, $r \rangle$ to $c$\;
}

\Event POdeliver($\langle$ $\delta$, $r$, $c$, $t\rangle$) \Do{
	$\Sigma \la apply(\delta, \Sigma)$\;
	set $\langle c, t \rangle$ as replied and store $r$ as last reply to $c$\;
	send $\langle c$, $t$, $r \rangle$ to $c$\;
}

\end{small}
\caption{Passive replication based on \POabcast\ - replica}
\label{alg:PB}
\end{algorithm}

\begin{algorithm}[t]

\begin{small}

\textbf{initially} \Do{
	$t \la 0$\; 
}

\Event execute operation $o$ \Do{
	$t \la t+1$\; 	
	reliably send $\langle c$, $t$, $o\rangle$ to all replicas\;
	wait for $\langle c$, $t$, $r\rangle$ from any replica\;
}

\end{small}

\caption{Passive replication based on \POabcast\ - client $c$}
\label{alg:PB-client}
\end{algorithm}

Primary order atomic broadcast (\POabcast) is an intermediate abstraction used by our unified passive replication algorithm.
\POabcast\ provides a broadcast primitive {\em POabcast} and a delivery primitive {\em POdeliver}.
\POabcast\ satisfies all safety properties of atomic broadcast.

\begin{defn}[Atomic broadcast]
An atomic broadcast primitive consists of two operations: {\em broadcast} and {\em deliver} of a value. 
It satisfies the following properties:
\begin{itemize}
\item[\textbf{\em Integrity}] If some process delivers $v$ then some process has broadcast $v$.
\item[\textbf{\em Total order}] If some process delivers $v$ before $v'$ then any process that delivers $v'$ must deliver $v$ before $v'$.
\item[\textbf{\em Agreement}] If some process $p_i$ delivers $v$ and some other process $p_j$ delivers $v'$, then either $p_i$ delivers $v'$ or $p_j$ delivers $v$.\footnote{We modified the traditional formulation of agreement to state it as a safety property only.}
\end{itemize}
\end{defn}

\POabcast\ extends atomic broadcast by introducing the concept of primary and a {\em barrier:} the additional {\em isPrimary}() primitive, which \POabcast\ uses to signal when a process is ready to broadcast state updates.
This predicate resembles $Prmy_s$ in the specification of Budhiraja {\em et al.}~\cite{primary-backup}.
However, as failure detectors are unreliable in our model, primary election is also unreliable: there might be multiple concurrent primaries at any given time, unlike in~\cite{primary-backup}.

 A {\em primary epoch} for a process $p$, or simply a {\em primary}, is a continuous period of time during which {\em isPrimary}() is true at $p$. 
Multiple primaries can be present at any given time: the {\em isPrimary}() predicate is local to a single process and multiple primary epochs can overlap in time.
Let $P$  be the set of primaries such that at least one value they propose is ever delivered by some process in the current run.
A {\em primary mapping} $\Lambda$ is a function that maps each primary in $P$ to a unique {\em primary identifier} $\lambda$, which we also use to denote the process executing the primary role.
We consider primaries as logical processes: saying that event $\epsilon$ occurs at primary $\lambda$ is equivalent to saying that $\epsilon$ occurs at some process $p$ during a primary epoch for $p$ having primary identifier $\lambda$.


\begin{defn}[Primary order atomic broadcast]
A primary order atomic broadcast primitive consists of two operations {\em broadcast}$(v)$ and {\em deliver}$(v)$, and of a binary {\em isPrimary}$()$ predicate, which indicates whether a process is a primary and is allowed to broadcast a value.
Let $\Lambda$ be a primary mapping and $\prec_\Lambda$ a total order relation among primary identifiers.
Primary order broadcast satisfies the Integrity, Total order, and Agreement properties of atomic broadcast; furthermore, it also satisfies the following additional properties:
\begin{itemize}
\item[\textbf{\em Local primary order}] If $\lambda$ broadcasts $v$ before $v'$, then a process that delivers $v'$ delivers $v$ before $v'$.
\item[\textbf{\em Global primary order}] If $\lambda$ broadcasts $v$, $\lambda'$ broadcasts $v'$,  $\lambda \prec_\Lambda \lambda'$, and some process $p$ delivers $v$ and $v'$, then $p$ delivers $v$ before $v'$.
\item[\textbf{\em Primary integrity}] If $\lambda$ broadcasts $v$, $\lambda'$ broadcasts $v'$, $\lambda \prec_\Lambda \lambda'$, and some process delivers $v$, then $\lambda'$ delivers $v$ before it broadcasts $v'$.
\end{itemize}
\end{defn}

These properties are partially overlapping, as we show in Appendix~\ref{sec:alternative-specs}. 
For example, global primary order is very useful in reasoning about the behaviour of \POabcast, but it can be implied from the other \POabcast\ properties.
It is also worth noting that local primary order is weaker than the single-sender FIFO property, since it only holds within a single primary epoch.

The above properties focus on safety. 
For liveness, it is sufficient to require the following:
\begin{defn}[Eventual Single Primary]
There exists a correct process such that eventually it is elected primary infinitely often and all messages it broadcasts are delivered by some process. 
\end{defn}

\begin{defn}[Delivery Liveness]
If a process delivers $v$ then eventually every correct process delivers $v$.
\end{defn}

\section{Passive Replication from \POabcast}
Before describing our unified \POabcast\ algorithm, we briefly show how to implement passive replication on top of \POabcast.

In passive replication systems, all replicas keep a copy of an object and receive operations from clients.
Each client waits for a reply to its last operation $op$ before it submits the following one.
Each operation has a unique identifier, comprising the client identifier $c$ and a counter $t$ local to the client.

The pseudocode of the passive replication algorithm is illustrated in Algorithms~\ref{alg:PB} and~\ref{alg:PB-client}.
This algorithm captures key aspects of practical systems, like the ZooKeeper coordination system.
In the case of ZooKeeper, for example, the object is the data tree used to store client information.

A primary replica is a process whose {\em isPrimary} predicate is true.
This predicate is determined by the underlying \POabcast\ algorithm.
We will see later in the paper that becoming primary entails crossing a {\em barrier}.

Replicas keep two states: a {\em committed state} $\Sigma$ and a {\em tentative state} $\Theta$.
The primary (tentatively) executes an operation $o$ on the tentative state $\Theta$; the primary remembers the identifier of the last executed operation from each client (modeled by the {\em replied}($c$, $t$) predicate) and the corresponding reply $r$.
The execution of a new operation generates a state update $\delta$, which the primary broadcasts to the other replicas together with the reply and the unique operation identifier $\langle c, t\rangle$.
We use $ S \tran{op} \langle r, \delta_{SQ} \rangle $ to denote that executing $op$ on the state $S$ produces a reply $r$ and a state update $\delta_{SQ}$, which determines a transition from $S$ to a new state $Q$.
When a replica delivers state updates, it applies them onto its committed state $\Sigma$.

We show in Appendix~\ref{sec:correct-passive-repl} that this algorithm implements a linearizable replicated object. 
We also discuss a counterexample showing why \POabcast\ cannot be replaced with atomic broadcast due to its lack of barriers;
it is a generalization of the counterexample discussed in~\cite{zab}.

\section{Unified \POabcast\ Algorithm Using the Barrier Function}

After motivating the use of \POabcast\ to obtain natural implementations of passive replication and discussing the role of the barrier predicate {\em isPrimary}, we introduce our unified $\barrier$-based \POabcast\ algorithm (Algorithm~\ref{alg:PB-barrier}).
It uses three underlying primitives: consensus, the $\Omega$ leader oracle, and a new {\em barrier function} $\barrier$ we will define shortly.

Like typical atomic broadcast algorithms, our \POabcast\ algorithm runs a sequence of consensus instances, each associated with an instance identifier~\cite{chandra-toueg-FD}.
Broadcast values are proposed using increasing consensus instance identifiers, tracked using the $prop$ counter.
Values are decided and delivered following the consensus instance order:
if the last decided instance was $dec$, only the event {\em decide}$(v, dec+1)$ can be activated, resulting in an invocation of {\em POdeliver}.
This abstracts the buffering of out-of-order decisions between the consensus primitive and our algorithm.

The most important difference between our algorithm and an implementation of atomic broadcast is that it imposes an additional barrier condition for broadcasting messages: it must hold {\em isPrimary}.
In particular, it is necessary for safety that $dec \geq \barrier$. 
The condition that a primary must be a leader according to $\Omega$ is only required for the liveness property of \POabcast: it ensures that eventually there is exactly one correct primary in the system.
The  barrier function $\barrier$ returns an integer and is defined as follows.


\begin{algorithm}[t]

\begin{small}

\textbf{initially} \Do{
$dec$ $\la$ 0\;
$prop$ $\la$ 0\;
}

\Event POabcast($v$) $\wedge$ isPrimary() \Do{
	$prop$ $\la$ $\max(prop + 1, dec + 1)$\;
	propose($v$, $prop$)\;
}

\Event decide($v$, $dec+1$) \Do{
	$dec \la dec + 1$\;
	POdeliver($v$)\;
}

\Funct isPrimary() \Do{
	\textbf{return} $(dec \geq \barrier)$ $\wedge$  $(\Omega = p$)\;
}

\vspace{0.5cm}

\tcc{Gap handling}

\Event $\Omega$ changes from $q \neq p$ to $p$ \Do{
	\ForAll{$i \in[dec+1,\tau]$}{
		propose($skip(\tau)$, $i$)\;
	}
}

\Event decide($skip(k)$, $dec+1$) \Do{
	$dec \la k$\;
}

\end{small}

\caption{\POabcast\ based on the barrier function and consensus - process $p$}
\label{alg:PB-barrier}

\end{algorithm}

\begin{defn}[Barrier function ]
\label{def:barrier}

Let $\sigma$ be an infinite execution, $\Lambda$ a primary mapping in $\sigma$, $\prec_{\Lambda}$ a total order among the primary identifiers,
and $\lambda$ a primary such that at least one value it proposes is delivered in $\sigma$.
A barrier function $\barrier$ for $\lambda$ returns:\footnote{Subscripts denote the process that executed the {\em propose} or {\em decide} steps.}

$$
\barrier = \max \{i: \exists v, p, \lambda' \: s.t. \: decide_p(v,i) \in \sigma \wedge propose_{\lambda'}(v,i) \in \sigma \wedge \lambda' \prec_{\Lambda} \lambda \}
$$
\end{defn}

%
%
%
%
%
%
%

%

An actual implementation of the $\barrier$ function can only observe the finite prefix of $\sigma$ preceding its invocation; however, it must make sure that its outputs are valid in any infinite extension of the current execution.
If none of the values proposed by a primary during a primary epoch are ever delivered, $\tau$ can return arbitrary values.

We show in Appendix~\ref{sec:correct-barrier} that this definition of $\tau$ is sufficient to guarantee the additional properties of \POabcast\ compared to atomic broadcast.
In particular, it is key to guarantee that the primary integrity property is respected.
Local primary order is obtained by delivering elements in the order in which they are proposed and decided.

The key to defining a barrier function is identifying a primary mapping $\Lambda$ and a total order of primary identifiers $\prec_{\Lambda}$ that satisfy the barrier property, as we will show in the following section.
There are two important observations to do here.
First, we use the same primary mapping $\Lambda$  and total order $\prec_{\Lambda}$  for the barrier function and for \POabcast.
If no value proposed by a primary $p$ is ever decided, then the primary has no identifier in $\Lambda$ and is not ordered by $\prec_{\Lambda}$, so the output of $\barrier$ returned to $p$ is not constrained by Definition~\ref{def:barrier}.
This is fine because values proposed by $p$ are irrelevant for the correctness of passive replication: they are never delivered in the \POabcast\ Algorithm~\ref{alg:PB-barrier} and therefore the corresponding state updates are never observed by clients in the passive replication Algorithms~\ref{alg:PB} and~\ref{alg:PB-client}.
Note also that a primary might not know its identifier $\lambda$: this is only needed for the correctness argument.

We call $\barrier$ a ``barrier" because of the way it is used in Alg.~\ref{alg:PB-barrier}. 
Intuitively, a new leader waits until it determines either that some value proposed by a previous primary is decided at some point in time or that no such value will ever be delivered.
Until the leader makes such a decision, it does not make further progress. 
Also, although this blocking mechanism is mainly implemented by the leader, the overall goal is to have all processes agreeing on the outcome for a value, so it resembles the classic notion of a barrier.

Consensus does not guarantee the termination of instances in which only a faulty primary has made a proposal. 
Therefore, a new primary proposes a special $skip(\tau)$ value to ensure progress for all instances in which no decision has been reached yet, that is, those with instance number between $dec+1$ and $\tau$.
If the skip value is decided, all decision events on instances up to $\tau$ are ignored, and therefore the values decided in these instances are not POdelivered.

\todo{Refer to TR instead of appendix. Add extended abstract for CR.}

\section{Implementations of the Barrier Function $\barrier$}
\label{sec:barrier-impl}


\subsection{Barriers with Black-Box Consensus}
\label{sec:black-box}

We first show how to implement $\barrier$ using the consensus primitive as a black box.
This solution is modular but imposes the use of sequential consensus instances: a primary is allowed to have at most one outstanding broadcast at a time.
This corresponds to the semi-passive replication algorithm~\cite{defago2004semi}.

Let $prop$ and $dec$ be the variables used in Algorithm~\ref{alg:PB-barrier}, and let $\tau_{seq}$ be equal to $\max(prop, dec)$.
We have the following result:

\begin{thm}
The function $\tau_{seq}$ is a barrier function.
\end{thm}

\noindent {\em Proof.}
We define $\Lambda$ as follows: if a leader process $p$ proposes a value $v_{i,p}$ for consensus instance $i$ and $v_{i,p}$ is decided, $p$ has primary identifier $\lambda = i$.
A primary has only one identifier: after $v_{i,p}$ is broadcast, it holds $prop>dec$ and $dec < \tau_{seq}$, so {\em isPrimary}() stops evaluating to true at $p$.
The order $\prec_{\Lambda}$ is defined by ordering primary identifiers as regular integers.

If a process $p$ proposes a value $v$ for instance $i = \max(prop+1, dec+1)$ in Algorithm~\ref{alg:PB-barrier}, it observes $\barrier_{seq} = \max(prop, dec) =  i-1$ when it becomes a primary.
If $v$ is decided, $p$ has primary identifier $\lambda = i$.
All primaries preceding $\lambda$ in $\prec_\Lambda$
have proposed values for instances preceding $i$, so $\tau_{seq}$ meets the requirements of barrier functions.$\hfill\Box$\bigskip

\subsection{Impossibility}
One might wonder if this limitation of sequential instances is inherent.
Indeed, this is the case as we now show.


\begin{thm}
Let $\Pi$ be a set of two or more processes executing the $\barrier$-based {\em \POabcast} algorithm with an underlying consensus implementation $C$ that can only be accessed through its propose and decide calls.
There is no local implementation of $\barrier$ for $C$ allowing a primary $p$ to propose a value for instance $i$ before $p$ reaches a decision for instance $i-1$.
\end{thm}

\noindent {\em Proof.}
The proof is by contradiction: we assume that a barrier function $\barrier_c$ allowing primaries to propose values for multiple concurrent consensus instances exists.

{\em Run $\sigma_1$:} The oracle $\Omega$ outputs some process $p$ as the only leader in the system from the beginning of the run.
Assume that $p$ broadcasts two values $v_1$ and $v_2$ at the beginning of the run.
For liveness of \POabcast, $p$ must eventually propose values for consensus instances $1$ and $2$.
By assumption, $\barrier_c$ allows $p$ to start consensus instance $2$ before a decision for instance $1$ is reached.
Therefore $p$ observes $\barrier_c = 0$ when it proposes $v_1$ and $v_2$.
The output of $\barrier_c$ must be independent from the internal events of the underlying consensus implementation $C$, since $\barrier_c$  cannot observe them.
We can therefore assume that no process receives any message 
before $p$ proposes $v_2$.

{\em Run $\sigma'_1$:} The prefix of $\sigma_1$ that finishes immediately after $p$ proposes $v_2$.
No process receives any message. 

{\em Run $\sigma_2$:} Similar to $\sigma_1$, but the only leader is $p' \neq p$ and the proposed values are $v'_1$ and $v'_2$.
Process $p'$ observes $\barrier_c = 0$ when it proposes $v'_1$ and $v'_2$.

{\em Run $\sigma'_2$:} The prefix of $\sigma_2$ that finishes immediately after $p'$ proposes $v'_2$.
No process receives any message. 

{\em Run $\sigma_3$:} The beginning of this run is the union of all events in the runs $\sigma'_1$ and $\sigma'_2$.
No process receives any message 
until the end of the union of $\sigma'_1$ and $\sigma'_2$.
The $\Omega$ oracle is allowed to elect two distinct leaders for a finite time.
Process $p$ (resp. $p'$) cannot distinguish between run $\sigma'_1$  (resp. $\sigma'_2$) and the corresponding local prefix of $\sigma_3$ based on the outputs of the consensus primitive and of the leader oracle.
After the events of $\sigma'_1$ and $\sigma'_2$ have occurred, some process decides $v'_1$ for consensus instance $1$ and $v_2$ for consensus instance $2$.

Regardless of the definition of $\Lambda$ and $\prec_\Lambda$, the output of $\barrier_c$ in $\sigma_3$ is incorrect.
Let $p$ and $p'$ have primary identifiers $\lambda$ and $\lambda'$ when they proposed $v_2$ and $v'_1$, respectively.
If $\lambda \prec_\Lambda \lambda'$, $\barrier_c$ should have returned $2$ instead of $0$ when $p'$ became primary.
If $\lambda' \prec_\Lambda \lambda$, $\barrier_c$ should have returned $1$ instead of $0$ when $p$ became primary.$\hfill\Box$\bigskip

\subsection{Barriers with White-Box Paxos}
\label{sec:white-box}

An alternative, corresponding to Zab~\cite{zab}, to avoid the aforementioned impossibility is to consider the internal states of the underlying consensus algorithm. 
We exemplify this approach considering the popular Paxos algorithm~\cite{paxos}. 
A detailed discussion of Paxos is out of the scope of this work and we only present a summary for completeness. 

\spara{Overview of Paxos.}
In Paxos each process keeps, for every consensus instance, an {\em accepted value}, which is the most current value it is aware of that might have been decided.
A process $p$ elected leader must first read, for each instance, the value that may have been decided upon for this instance, if any.
To obtain this value, the leader selects a unique {\em ballot number} $b$ and executes a {\em read phase} by sending a read message to all other processes.
Processes that have not yet received messages from a leader with a higher ballot number $b$ reply by sending their current accepted value for the instance.
Each accepted value is sent attached to the ballot number of the previous leader that proposed that value.
The other processes also promise not to accept any message from leaders with ballots lower than $b$.
When $p$ receives accepted values from a majority of processes, it {\em picks} for each instance the accepted value with the highest attached ballot.

After completing the read phase, the new leader proposes the values it picked as well as its own values for the instances for which no value was decided.
The leader proposes values in a {\em write} phase: it sends them to all processes together with the current ballot number $b$.
Processes accept proposed values only if they have not already received messages from a leader with a ballot number $b' > b$.
After they accept a proposed value, they send an acknowledgement to the leader proposing it.
When a value has been written with the same ballot at a majority of processes, it is decided.

In a nutshell, the correctness argument of Paxos boils down to the following argument.
If a value $v$ has been decided, a majority of processes have accepted it with a given ballot number $b$; we say that the proposal $\langle v, b \rangle$ is {\em chosen}.
If the proposal is chosen, no process in the majority will accept a value from a leader with a ballot number lower than $b$.
At the same time, every leader with a ballot number higher than $b$ will read the chosen proposal in the read phase, and will also propose the $v$.

\spara{Integrating the Barrier Function.}
We modify Paxos to incorporate the barrier function.
If a process is not a leader, there is no reason for evaluating $\barrier$.
Whenever a process is elected leader, it executes the read phase.
Given a process $p$ such that $\Omega = p$, let $read(p)$ be the maximum consensus instance for which any value has been picked in the last read phase executed by $p$.
The barrier function is implemented as follows:

$$
\begin{array}{l}
\barrier_{\textrm{Paxos}} = \left\{ \begin{array}{ll}
					\top & \textrm{ iff } \: \Omega \neq p \; \vee p \textrm{ is in read phase}\\
					read(p) & \textrm{ iff } \: \Omega = p \; \wedge p \textrm{ is in write phase }
\end{array} \right.
\end{array}
$$

The output value $\top$ is such that $dec \geq \barrier_{Paxos}$ never holds for any value of $dec$.
This prevents leaders from becoming primaries until a correct output for $\barrier_{Paxos}$ is determined.

The sequence of values picked by the new leader $p$ can have gaps. This occurs when $p$ can pick some value for an instance $i$ but it cannot pick any value for an instance $j < i$.
The new leader fills such gaps with {\em no-op} values.
This is analogous to the gap handling mechanism in Algorithm~\ref{alg:PB-barrier}, where the new primary proposes $skip(\tau)$ values to fill gaps. 
Therefore, whenever Paxos decides a value {\em no-op} for an instance $i$, Algorithm~\ref{alg:PB-barrier} treats this decision as if consensus decided $skip(\tau)$ for $i$.

%
%
%

We now show that this $\barrier$ implementation is correct.
The proof relies on the correctness argument of Paxos.

\begin{thm}
The function $\tau_{\textrm{Paxos}}$ is a barrier function.
\end{thm}

\noindent {\em Proof.} 
By the definition of $\barrier_{Paxos}$, a process becomes a primary if it is a leader and has completed the read phase. 
Let $\Lambda$ associate a primary  with the unique ballot number it uses in the Paxos read phase if some values it proposes is ever decided,
and let $\prec_{\Lambda}$ be the ballot number order.
Paxos guarantees that if any process ever decides a value $v$ proposed by a leader with ballot number smaller than the one of $\lambda$, then $v$ is picked by $\lambda$ in the read phase~\cite{paxos}.
This is sufficient to meet the requirements of $\barrier$.$\hfill\Box$\bigskip


\subsection{Time Complexity of $\barrier$-Based \POabcast\ with Different Barrier Functions} 
\label{sec:barrier-compl}

We now explain the second and third row of Table~\ref{tab:comp}. 
Just for the analysis, we assume that there are $c$ clients in the system, that the communication delay is $\Delta$, and that Paxos is used as underlying consensus protocol in all our implementations since it is optimal~\cite{lower}.

We first consider the barrier function of Sect.~\ref{sec:black-box}.
If a primary receives requests from all clients at the same time, it will broadcast and deliver the corresponding state updates sequentially.
Delivering a message requires $2\Delta$, the latency of the write phase of Paxos. 
Since each message will take $2\Delta$ time to be delivered, the last message will be delivered in $2\Delta \cdot c$ time.
During leader change, Paxos takes $2\Delta$ time to execute the read phase and $2\Delta$ to execute the write phase if a proposal by the old primary has been chosen and potentially decided in the last consensus instance.

With the barrier function of Sect.~\ref{sec:white-box}, consensus instances are executed in parallel with a latency of $2 \Delta$. 
The complexity for leader changes is the same, since the write phase is executed in parallel for all instances up to $\tau$.

Note that the longer leader change time of \POabcast\ algorithms compared to atomic broadcast (see Table~\ref{tab:comp}) is due to the barrier: before it becomes a primary, a process must {\em decide } on all values that have been proposed by the previous primaries and potentially decided (chosen). This is equivalent to executing read and write phases that require $4\Delta$ time.
In atomic broadcast, it is sufficient that a new leader {\em proposes} chosen values from previous leaders.

\subsection{Relationship between $\barrier$ Functions and Existing \POabcast\ Algorithms}
\label{sec:existing}

The \POabcast\ algorithm with the barrier function of Sect.~\ref{sec:black-box} is similar to semi-passive replication~\cite{defago2004semi} since both enforce the same constraint: primaries only keep one outstanding consensus instance at a time.
The time complexity of the two protocols using Paxos as the underlying consensus protocol is the same (Table~\ref{tab:comp}, second row).

If the barrier function implementation selects a specific consensus protocol and assumes that it can access its internal state, as discussed in Sect.~\ref{sec:black-box}, our barrier-based \POabcast\ algorithm can broadcast state updates in the presence of multiple outstanding consensus instances.
This is the same approach as Zab, and indeed there are many parallelisms with this algorithm.
The time complexity in stable periods is the same (see Table~\ref{tab:comp}, third row).
A closer look shows that also the leader change complexity is equal, apart from specific optimizations of the Zab protocol.
In Zab, the read phase of Paxos corresponds to the {\em discovery phase}; the CEPOCH message is used to implement leader election and to speed up the selection of a unique ballot (or epoch, in Zab terms) number that is higher than any previous epoch numbers~\cite{zab}.
After the read phase is completed, the leader decides on all consensus instances until the instance identifier returned by $\barrier_{Paxos}$ - this is the {\em synchronization phase}, which corresponds to a write phase in Paxos; in our implementation, the barrier function returns and the leader waits until enough consensus instances are decided.
At this point, the necessary condition $dec \geq \barrier_{Paxos}$ of our generic \POabcast\ construction is fulfilled, so the leader crosses the barrier, becomes a primary, and can proceed with proposing values for new instances.
In Zab, this corresponds to the {\em broadcast phase}.

Virtually-synchronous Paxos is also a modified version of Paxos that implements \POabcast\ and the $\barrier_{Paxos}$ barrier function, but it has the additional property of making the set of participating processes dynamic~\cite{vs-paxos}.
It has the same time complexity during stable periods and leader changes as in Table~\ref{tab:comp}.

\section{\POabcast\ Using Consensus Instead of $\barrier$ for the Barrier}
\label{sec:simulation}
The previous section shows an inherent tradeoff in $\barrier$ implementations between modularity, which can be achieved by using sequential consensus instances and using consensus as a black box, and performance, which can be increased by integrating the implementation of the barrier function in a specific consensus protocol.
In this section, we show that this tradeoff can be avoided through the use of an alternative \POabcast\ algorithm.

\newcommand{\tentativeepoch}{{\em tent-epoch}}
\newcommand{\currentepoch}{{\em epoch}}
\newcommand{\msgtentativeepoch}{{\em tent-epoch$_m$}}
\newcommand{\decarray}{{\em dec-array}}
\newcommand{\proparray}{{\em prop-array}}
\newcommand{\comm}{{\em deliv-seqno}}
\newcommand{\seqno}{{\em seqno}}
\newcommand{\val}{ {\em v}}
\newcommand{\msgepoch}{{\em epoch$_m$}}
\newcommand{\msgseqno}{{\em seqno$_m$}}
\newcommand{\dec}{{\em dec}}
\newcommand{\proposal}{{\em prop}}
\newcommand{\ready}{{\em primary}}

\subsection{Algorithm}
Our $\barrier$-free algorithm (see Algorithm~\ref{alg:simul}) implements \POabcast, so it is an alternative to Algorithm~\ref{alg:PB-barrier}.
The algorithm is built upon a leader election oracle $\Omega$ and consensus.
The main difference with Algorithm~\ref{alg:PB-barrier} is that the barrier predicate {\em isPrimary} is implemented using consensus instead of $\barrier$: consensus instances are used to agree not only on values, but also on primary election information.
Another difference is that some decided value may not be delivered.
This requires the use of additional buffering, which slightly increases the complexity of the implementation.

When a process $p$ becomes leader, it picks a unique epoch number \tentativeepoch\ and proposes a $\langle$\newepoch,~\tentativeepoch$\rangle$ value in the smallest consensus instance \dec\ in which $p$ has not yet reached a decision (lines~\ref{ln:try-primary-start}-\ref{ln:try-primary-end}).
Like in Algorithm~\ref{alg:PB-barrier}, we use multiple consensus instances.
All replicas keep a decision counter \dec, which indicates the current instance in which a consensus decision is  awaited, and a proposal counter \proposal, which indicates the next available instance for proposing a value.
Another similarity with Algorithm~\ref{alg:PB-barrier} is that decision events are processed following the order of consensus instances, tracked using the variable \dec\ (see lines~\ref{ln:new-epoch-start} and~\ref{ln:val-start}).
Out-of-order decision events are buffered, although this is omitted in the pseudocode.

Every time a \newepoch\ tuple is decided, the sender of the message is elected primary and its epoch \tentativeepoch\ is {\em established} (lines~\ref{ln:new-epoch-start}-\ref{ln:new-epoch-end}).
When a new epoch is established, processes set their current epoch counter \currentepoch\ to \tentativeepoch.
If the process delivering the \newepoch\ tuple is a leader, it checks whether the epoch that has been just established is its own tentative epoch.
If this is the case, the process considers itself as a primary and sets \ready\ to true;
else, it tries to become a primary again.


\begin{algorithm*}[t]

\RestyleAlgo{plain}
\LinesNumbered

\begin{minipage}[t]{0.45\linewidth}
\begin{algorithm}[H]
\begin{small}

\textbf{initially} \Do{
\tentativeepoch, \dec, \comm, \proposal, \seqno\ $\la 0$\;
\currentepoch\ $\la \bot$\;
\ready\ $\la$ false\;
}

\Event $\Omega$ changes from $q \neq p$ to $p$ \nllabel{ln:try-primary-start} \Do{
    try-primary()\;
}

\Proc try-primary() \Do{
    \tentativeepoch\ $\la$ new unique epoch number\;
    propose($\langle$\newepoch, \tentativeepoch$\rangle$, \dec)\nllabel{ln:try-primary-end}\;
}

\Event decide($\langle$\newepoch, \msgtentativeepoch $\rangle$, \dec)\nllabel{ln:new-epoch-start} \Do{
    \dec\ $\la$ \dec$+1$\nllabel{ln:epoch-reset-start}\;
    \currentepoch\ $\la$ \msgtentativeepoch\;
    \decarray\ $\la$ empty array\;
    \proparray\ $\la$ empty array\;
    \comm\ $\la$ \dec\nllabel{ln:epoch-reset-end}\;
    \If{$\Omega = p$}{
        \If{\em \tentativeepoch\ $=$ \msgtentativeepoch}{
            \proposal\ $\la$ \dec\;
            \seqno\ $\la$ \dec\;
            \ready\ $\la$ true\;
        }
        \Else{
            \ready\ $\la$ false\;
            try-primary()\nllabel{ln:new-epoch-end}\;
        }
    }
}


\end{small}
\end{algorithm}
\end{minipage}
\hspace{1mm}
\RestyleAlgo{plain}
\LinesNumbered
\begin{minipage}[t]{0.45\linewidth}

\begin{algorithm}[H]
\begin{small}

\Event POabcast(\val)\nllabel{ln:abcast-start}\Do{
    propose($\langle$\msgvalue, \val, \currentepoch, \seqno$\rangle$, \proposal)\;
    \proparray[\proposal] $\la \langle$\val, \seqno $\rangle$\;
    \proposal $\la$ \proposal $+ 1$\;
    \seqno\ $\la$ \seqno $+ 1$\nllabel{ln:abcast-end}\;
}

\Event decide($\langle$\msgvalue, \val, \msgepoch, \msgseqno$\rangle$, \dec)\nllabel{ln:val-start} \Do{
    \If{\em \msgepoch\ $=$ \currentepoch}{
        \decarray[\msgseqno] $\la$ \val\;
        \While{\em \decarray$[$\comm$]$ $\neq \bot$}{\nllabel{ln:decarray}
            POdeliver(\decarray[\comm])\;
            \comm $\la$ \comm $+1$\;\nllabel{ln:deliver-end}
        }
    }
    \If{\em \ready\ $\wedge$ \msgepoch $\neq$ \currentepoch\ \\ $\wedge$ \proposal\ $\geq$ \dec}{
        $\langle$\val$'$, \seqno$'\rangle \la$ \proparray[\dec]\;
        \proparray[\proposal] $\la$ \proparray[\dec]\;
        propose($\langle$\msgvalue, \val$'$, \currentepoch, \seqno$'\rangle$, \proposal)\;
        \proposal $\la$ \proposal $+1$\;
    }
    \If{\em $\neg$ \ready\ $\wedge \: \Omega = p$}{
        try-primary()\;
    }
    \dec $\la$ \dec $+1$\;
}\nllabel{ln:val-end}

\Event $\Omega$ changes from $p$ to $q \neq p$ \Do{
    \ready $\la$ false\;
}

\Funct isPrimary() \Do{
    \textbf{return}  \ready\;
}

\end{small}

\end{algorithm}
\end{minipage}

\caption{Barrier-free \POabcast\ using black-box consensus - process $p$}
\label{alg:simul}
\end{algorithm*}

When $p$ becomes a primary, it can start to broadcast values by proposing \msgvalue\ tuples in the next consensus instances, in parallel (lines~24-28).
Ensuring that followers are in a state consistent with the new primary does not require using barriers: all processes establishing  \tentativeepoch\ in consensus instance $i$ have decided and delivered the same sequence of values in the instances preceding~$i$.
This guarantees that the primary integrity property of \POabcast\ is respected.

Processes only POdeliver \msgvalue\ tuples of the last established epoch until a different epoch is established (lines~\ref{ln:val-start}-33, see in particular condition \msgepoch\ $=$ \currentepoch).
The algorithm establishes the following total order $\prec_\Lambda$ of primary identifiers: given two different primaries $\lambda$ and $\lambda'$ which picked epoch numbers $e$ and $e'$ respectively, we say that $\lambda \prec_\Lambda \lambda'$ if and only if a tuple $\langle$\newepoch, $e\rangle$ is decided for a consensus instance $n$, a tuple $\langle$\newepoch, $e'\rangle$ is decided for a consensus instance $m$, and $n < m$.
Suppose that $p$ is the primary $\lambda$ with epoch number $e_\lambda$ elected in consensus instance \dec$_\lambda$.
All processes set their current  epoch variable $e$ to $e_\lambda$ after deciding in instance \dec$_\lambda$.
From consensus instance number \dec$_\lambda+1$ to the next consensus instance in which a \newepoch\ tuple is decided, processes decide and deliver only values that are sent from $\lambda$ and included in \msgvalue\ tuples with \msgepoch\ $ = e_\lambda$.
Replicas thus deliver messages following the order $\prec_\Lambda$ of the primaries that sent them, fulfilling the global primary order property of \POabcast.

The additional complexity in handling \msgvalue\ tuples is necessary to guarantee the local primary order property of \POabcast.
\msgvalue\ tuples of an epoch are not necessarily decided in the same order as they are proposed.
This is why primaries include a sequence number \seqno\ in \msgvalue\ tuples.
In some consensus instance, the tuples proposed by the current primary might not be the ones decided.
This can happen in the presence of concurrent primaries, since primaries send proposals for multiple overlapping consensus instances without waiting for decisions.
If a primary is demoted, values from old and new primaries could be interleaved in the sequence of decided values for a finite number of instances.
All processes agree on the current epoch of every instance, so they do not deliver messages from other primaries with different epoch numbers.
However, it is necessary to buffer out-of-order values from the current primary to deliver them later.
That is why processes store decided values from the current primary in the \decarray\ array (line~31), and deliver them only if a continuous sequence of sequence numbers, tracked by \comm, can be delivered (lines~32-34).

Primaries also need to resend \msgvalue\ tuples that could not be decided in the correct order.
When values are proposed, they are stored in the \proparray\ following the sequence number order; this buffer is reset to the next ongoing consensus instance every time a new primary is elected.
Primaries resend \msgvalue\ tuples in lines~35-40.
Primaries keep a proposal instance counter \proposal, indicating the next consensus instance in which values can be proposed.
If an established primary has outstanding proposals for the currently decided instance \dec, it holds \proposal\ $\geq$ \dec.
In this case, if the decided \msgvalue\ tuple is not one such outstanding proposal but has instead been sent by a previous primary, it holds that \msgepoch\ $\neq$ \currentepoch.
If all the previous conditions hold, the established primary must resend the value that has been skipped, \proparray$[$\dec$].v'$, using the same original sequence number \proparray$[$\dec$].$\seqno' in the next available consensus instance, which is \proposal.

The arrays \decarray\ and \proparray\ do not need to grow indefinitely.
Elements of \decarray\ (resp.~\proparray) with position smaller than \comm\ (resp.~\dec) can be garbage-collected.

For liveness, a leader which is not a primary keeps trying to become a primary by sending a \newepoch\ tuple for every consensus instance (lines~22-23).
The \ready\ variable is true if a leader is an established primary.
It stops being true if the primary is not a leader any longer (lines~44-45).

%
%
%
%
%

Algorithm~\ref{alg:simul} correctly implements \POabcast, as shown in the Appendix~\ref{sec:correct-barrier-free}.

\subsection{Time Complexity}
\label{sec:barrier-free-compl}

As before, we use Paxos for the consensus algorithm and assume a communication delay of $\Delta$.
During stable periods, the time to deliver a value is $2\Delta$, which is the time needed to execute a Paxos write phase.
When a new leader is elected, it first executes the read phase, which takes $2\Delta$.
Next, it executes the write phase for all instances in which values have been read but not yet decided, and for one additional instance for its \newepoch\ tuple.
All these instances are executed in parallel, so they finish within $2\Delta$ time.
After this time, the new leader crosses the barrier, becomes a primary, and starts broadcasting new values.

\begin{figure}[t]
    \centering
    \includegraphics[width=0.45\linewidth]{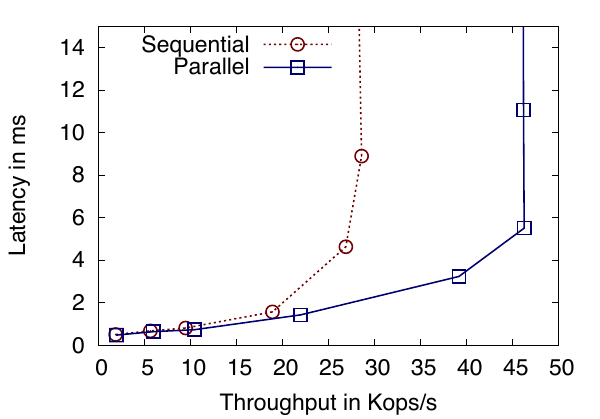} 
    \caption{Latency and throughput with micro benchmarks. Request and state update sizes were set to  1kb, which is the typical size observed in ZooKeeper. Both protocols use batching}
    \label{fig:evaluation}
\end{figure}

\section{Experimental Evaluation}
\label{sec:evaluation}
%

Our $\barrier$-free algorithm combines modularity with constant time complexity.
Since our work was motivated by our work on systems like ZooKeeper, one might wonder whether this improvement has a practical impact.
Current implementations of replicated systems can reach some degree of parallelism even if they execute consensus instances sequentially. 
This is achieved through an optimization called {\em batching}: multiple clients requests are aggregated in a batch and agreed upon together using a single instance.
Even in presence of batching, we found that there is a substantial advantage of running multiple consensus instances in parallel.

We implemented two variants of the Paxos algorithm, one with sequential consensus instances and one with parallel ones, and measured the performance of running our \POabcast\ algorithms on top of it.
We consider fault-free runs in which the leader election oracle outputs the same leader to all processes from the beginning.
We used three replicas and additional dedicated machines for the clients; all servers are quad-core 2.5 GHz CPU servers with 16 GB of RAM connected through a Gigabit network.

The experiments consist of micro-benchmarks, in which the replicated object does nothing and has an empty state.
These benchmarks are commonly used in the evaluation of replication algorithms because they reproduce a scenario in which the replication protocol, rather than execution, is the bottleneck of the system. 
In systems where execution is the bottleneck, using a more efficient replication algorithm has no impact on the performance of the system.

We used batching in all our experiments.
With sequential consensus instances, we batch all requests received while a previous instance is ongoing. 
In the pipelined version, we start a new consensus instance when either the previous instance is completed or $b$ requests have been batched.
We found $b=50$ to be optimal.
Every measurement was repeated five times at steady state, and variances were negligible.

Figure~\ref{fig:evaluation} reports the performance of the two variants with messages (requests and state updates) of size 1 kB, which is a common state update size for ZooKeeper and Zab~\cite{zab}.
Each point in the plot corresponds of a different number of clients concurrently issuing requests.

The peak throughput with the parallel consensus instances is almost two times the one with sequential instances.
The same holds with messages of size 4 kB.
The difference decreases with smaller updates than the ones we observe in practical systems like ZooKeeper.
In the extreme case of empty requests and state updates, the two approaches have virtually the same request latency and throughput: they both achieve a maximum throughput of more than 110 kops/sec and a minimum latency of less than 0.5 ms. 

These results show that low time complexity (see Table~\ref{tab:comp}) is very important for high-performance passive replication. 
When there is little load in the system, the difference in latency between the two variants is negligible.
In fact, due to the use of batching, running parallel consensus instances is not needed.
Since clients can only submit one request at a time, we increase load by increasing the number of clients concurrently submitting requests;
this is equivalent to increasing the parameter $c$ of Table~\ref{tab:comp}.
As the number of clients increase, latency grows faster in the sequential case, as predicted by our analysis.
With sequential consensus instances, a larger latency also results in significantly worse throughput compared to the parallel variant due to lower network and CPU utilization.

\section{Conclusions}

Some popular systems such as ZooKeeper have used passive replication to mask crash faults. 
We have shown that the barrier predicate {\em isPrimary} is a key element differentiating active and passive replication, and its implementation significantly impacts the complexity and modularity of the algorithm.
We have shown how to implement passive replication on top of \POabcast.
By making leader processes cross the barrier before becoming primaries, we prevent state updates from being decided and applied out of order. 

We then extracted a unified algorithm for implementing \POabcast\ using the barrier function that abstract existing approaches.
The barrier function is a simple way to understand the difference between passive and active replication, as well as the characteristics of existing \POabcast\ algorithms, but it imposes a tradeoff between parallelism and modularity. 
We have proposed an algorithm that does present such a limitation by not relying upon a barrier function and yet it guarantees the order of state updates according to the succession of primaries over time.
This algorithm is different from existing ones in its use of consensus, instead of barrier functions, for primary election.  

\subsubsection*{Acknowledgement}
We would like to express our gratitude to Alex Shraer and Benjamin Reed for the insightful feedback on previous versions of the paper, and to Daniel G\'{o}mez Ferro for helping out with the experiments.

\bibliography{DISC-TR}
\bibliographystyle{plain}

\newpage
\appendix

\section{Alternative Definitions of \POabcast}
\label{sec:alternative-specs}

There are alternative formulation of \POabcast.
For example, while global primary order is important for understanding the properties of \POabcast, it can be implied by the other properties of \POabcast.

\begin{thm} {\em Global primary order can be implied from the other properties of \POabcast. }
\end{thm}

\noindent {\em Proof.} Assume that a process $p_i$ delivers two values $v$ and $v'$, which were broadcast by the primaries $\lambda$ and $\lambda'$ respectively.
Also, assume that $\lambda \prec_{\Lambda} \lambda'$.
Global primary order requires that $p_i$ delivers first $v$ and then $v'$.
Assume by contradiction that $p_i$ delivers the two values in the opposite order.
By primary integrity, $\lambda'$ delivers $v$ before broadcasting $v'$.
By total order, $\lambda'$ delivers $v'$ before $v$.
However, by integrity, $\lambda'$ cannot deliver $v'$ since it did not broadcast it yet, a contradiction.$\hfill\Box$

\section{Correctness of Passive Replication on Top of \POabcast}
\label{sec:correct-passive-repl}

This section shows the correctness of Algorithm~\ref{alg:PB}.

\spara{Formalism and rationale.} 
Replicas keep a copy of a shared object.
The shared object can be nondeterministic: there may be multiple state updates $\delta_{SQ_1}, \delta_{SQ_2}, \ldots$ and replies $r_1, r_2, \ldots$ such that $ S \tran{op} \langle \delta_{SQ_i}, r_i \rangle$.
We omit subscripts from state updates when these are not needed.

Applying state updates responds to the constraints we discussed in the introduction: 
state updates can only be applied on the state from which they were generated.
Formally, given a state update $\delta_{SQ}$ such that $ S \tran{op} \langle r, \delta_{SQ} \rangle $ for any value of $r$, we have the following.
$$
\begin{array}{l}
apply(\delta_{SQ}, R) = \left\{ \begin{array}{ll}
					Q \: \textrm{ iff } \: S = R\\
					\bot \: \textrm{ otherwise}
\end{array} \right.
\end{array}
$$
We also use the notation $S \Tran{op} Q$ to denote the tentative execution of the operation $op$ on a tentative state $S$ resulting in a transition to a tentative state $Q$.

%
%
%
%
%
%
%
%

The notation $\bot$ denotes an undefined state.
A correct passive replication system shall never reach such undefined state.
This formalizes the difference between agreeing on state updates and operations.
The execution of operations never leads the system to an undefined state. 
The order of their execution may be constrained by consistency requirements, for example linearizability, but not by the semantics of the replicated object.
State updates are different: they need to be applied exactly on the state in which they were generated or an undefined state $\bot$ can be reached. 

\begin{figure}[b]
    \centering
    \includegraphics[width=0.5\linewidth]{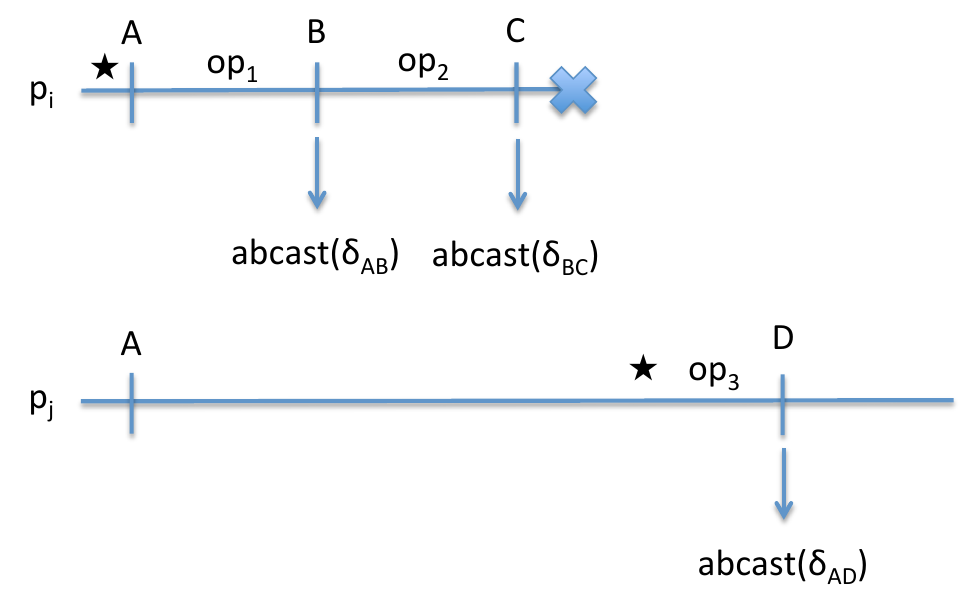} 
    \caption{In this run, atomic broadcast allows delivering the incorrect sequence of state updates $\langle \delta_{AD},\delta_{BC}\rangle$. The star denotes becoming a primary. $A,B,C,D$ are tentative states.}
    \label{fig:counter}
\end{figure}

The reason for adopting \POabcast\ instead of regular atomic broadcast in Algorithm~\ref{alg:PB} can be illustrated with an example (see Fig.~\ref{fig:counter}).
Consider an algorithm similar to Algorithm~\ref{alg:PB} such that atomic broadcast is used instead of \POabcast.
Assume also that {\em isPrimary} is implemented using reliable leader election, that is, only one primary process exists from the beginning, and it is a correct process. 
Consider a run in which the initial state of all processes is $A$.
A process $p_i$ is elected leader; it tentatively executes two operations $op_1$ and $op_2$ such that $A \Tran{op_1} B \Tran{op_2} C$, and it atomically broadcasts state updates $\delta_{AB}$ and $\delta_{BC}$.
Assume now that $p_i$ crashes and $p_j$ becomes leader. 
$p_j$ did not deliver any message so far, so it is still in state $A$. 
Process $p_j$ executes operation $op_3$ such that $A \Tran{op_3} D$, and broadcasts $\delta_{AD}$.

In this run there are no concurrent primaries.
Yet, atomic broadcast leads to an incorrect behavior since it could deliver the sequence of state updates $\langle \delta_{AD},\delta_{BC}\rangle$.
Since all delivered state updates are executed, a backup would execute $apply(\delta_{AD}, A)$, transitioning to state $D$, and then execute $apply(\delta_{BC}, D)$, transitioning to an undefined state $\bot$ that may not be reachable in correct runs.
Note that even using FIFO atomic broadcast would not solve the problem.
It guarantees that if a message from a process is delivered then all messages previously broadcast from that process are previously delivered.
In our example, FIFO atomic broadcast could still deliver the sequence of state updates $\langle \delta_{AD},\delta_{AB},\delta_{BC}\rangle$, which also leads to $\bot$.

\POabcast\ prevents this incorrect run. 
First, it determines a total order between primary identifiers: for example, this could order the primary identifier of $p_i$ before the one of $p_j$.
Global primary order ensures that followers apply state updates following this order: the updates sent by $p_j$ are delivered after the ones sent by $p_i$.
Primary integrity guarantees that, before broadcasting a state update, a primary delivers all state updates from previous primaries, thus making sure that all followers can safely apply the state updates it broadcasts.
For example, $p_j$ would apply the sequence of state updates $\langle \delta_{AB}, \delta_{BC} \rangle$, reaching state $C$ before it executes $op_3$ and starts sending state updates: therefore, it will not broadcast the state update $\delta_{AC}$.

\spara{Correctness.} The following is a more general correctness result.

\begin{thm} 
\label{lem:PB}
Passive replication based on primary order atomic broadcast satisfies linearizability.
\end{thm}

\noindent {\em Proof.}
Since \POabcast\ ensures the agreement and total order properties of atomic broadcast, it delivers state updates in a consistent total order $\prec_\Delta$.
We only need to consider delivered state updates since they are the only ones whose effects are visible to clients.

Consider a state update $\delta = \delta_{SQ}$ that is delivered by a process $p$ and has position $i$ in the sequence of state updates delivered by $p$.
Let $\delta$ be broadcast by some primary $\lambda$ that has generated $\delta_{SQ}$ by tentatively executing some operation $op$ on its local tentative state $\Theta = S$. 
We show that, for any value of $i \geq 0$ that: 
\begin{itemize}
\item[(i)] No state update preceding $\delta$ in $\prec_\Delta$ is generated by tentatively executing $op$;
\item[(ii)] $S$ is the state $p$ reaches after delivering the state updates preceding $\delta$ in $\prec_\Delta$.
\end{itemize}
Predicate (i) establishes a mapping of delivered state updates to operations; therefore, $\prec_\Delta$ can be used to determine an execution order $\prec_o$ of operations.
Predicate (ii) shows that replies to clients are legal and consistent with $\prec_o$.
If both predicates hold, then replicas transition from the state $S$ to the state $Q$ upon delivering the state update $\delta_{SQ}$.
Since $S \tran{op} \langle Q, r \rangle$, $Q$ is a legal state and $r$ a legal reply.
Therefore, linearizability follows by showing these two properties for any value of $i \geq 0$.

Both properties (i) and (ii) trivially hold for $i=0$ since all processes start from the same initial state.
We now consider the case $i > 0$.

Let $\delta'$ be the last state update delivered by $\lambda$ before POabcasting any value.
It follows from the integrity property of atomic broadcast that $\delta' \prec_\Delta \delta$.
Let $k < i$ be the position of $\delta'$ in the sequence of operations delivered by $\lambda$.
We use the notation $(\delta', \delta)_{\prec_\Delta}$ to indicate all operations delivered by $\lambda$ between $\delta'$ and $\delta$, which have positions in $(k,i)$.

We now consider two cases.


{\em Case 1: All state updates in $(\delta', \delta)_{\prec_\Delta}$ are broadcast by $\lambda$. }
In our algorithm, when \POabcast\ delivers a state update, it also delivers the identifier of the request that was executed to obtain the update.
Therefore, $\lambda$ knows the identifier of all requests related to all state updates preceding and including $\delta'$ in the delivery order $\prec_\Delta$, and consider them as replied.
Primaries generate state updates only for operations that have not been replied, so $op$ is not executed to generate any state updates up to and including $\delta'$ in the delivery order $\prec_\Delta$.
The primary $\lambda$ also knows the identifiers of all the operations it executed to generate the state updates in $(\delta',\delta]_{\prec_\Delta}$.
It executes $op$ and broadcasts $\delta$ only if it knows that $op$ was not already executed, ensuring property (i).

For property (ii), both $p$ and $\lambda$ deliver and apply the same sequence of state updates preceding and including $\delta'$ in $\prec_\Delta$.
It follows from Local Primary Order that $p$ commits all state updates in $(\delta',\delta]_{\prec_\Delta}$ in the same order as they are generated by $\lambda$.

{\em Case 2: There exists a state update in $(\delta', \delta)_{\prec_\Delta}$ that has been broadcast by another primary $\lambda' \neq \lambda$. }
This case leads to a contradiction.
If $\lambda \prec_\Lambda \lambda'$, a contradiction follows from global primary order, which requires that $p$ deliver $\delta$ before $\delta'$.
If $\lambda' \prec_\Lambda \lambda$, a contradiction follows from primary integrity: since $p$ delivers $\delta''$ before $\delta$, $\lambda$ must deliver $\delta''$ before $\delta'$, so $\delta'' \prec_\Delta \delta'$. $\hfill\Box$\bigskip

Liveness directly follows from the liveness property of \POabcast.
All requests from correct clients eventually complete, since correct clients resend their requests to the current primary.
%

\section{Correctness of the Unified \POabcast\ Algorithm}
\label{sec:correct-barrier}

We now show the correctness of Algorithm~\ref{alg:PB-barrier}.

\begin{thm} Primary order atomic broadcast based on the barrier function is correct.
\label{lem:poabcast}
\end{thm}

\noindent {\em Proof.} Processes exchange values using consensus instances; this ensures that all atomic broadcast properties are respected.
Local primary order follows by three observations. 
First, if a process proposes a value for an instance $i$, it will never propose a value for an instance $j < i$ since $prop$ only increases.
Second, values are decided and delivered according to the order of consensus instance identifiers.
Three, skip values do not determine gaps in the sequence of values proposed by a primary.

For primary integrity, consider that in order to broadcast a value, a process needs to become primary first. 
Let $\lambda$ be the primary identifier of a process.
By definition of barrier function, the last invocation of $\barrier$ returns to $\lambda$ a value $i$ greater or equal than the maximum identifier of a consensus instance in which a value proposed by any previous primary $\lambda' \prec_{\Lambda} \lambda$ is decided. 
Before broadcasting any message, $\lambda$ decides on all instances up to $i$ and delivers all values proposed by previous primaries that are delivered.
If $\lambda$ proposes a skip value and this is decided for an instance number $j < i$, no value from previous primaries decided between $j$ and $i$ is delivered.

Global primary order follows from the other properties of \POabcast.

The liveness properties of the algorithm follow from the Termination property of the underlying consensus primitive and from the Leader election oracle property of $\Omega$.
If some previous primary is faulty, the consensus instances it initiated may not terminate.
In this case, a new correct primary proposes skip values to guarantee progress.
Since the new primary proposes values starting from instance $\barrier$, skipping to $\barrier$ is sufficient to guarantee liveness when eventually a correct primary is elected.
$\hfill\Box$\bigskip

\section{Correctness of the Barrier-Free Algorithm}
\label{sec:correct-barrier-free}

This section shows that Algorithm~\ref{alg:simul} correctly implements \POabcast.

\begin{defn}[Total order of primaries]
Given two primaries $\lambda$ and $\lambda'$ with epoch numbers $e$ and $e'$ respectively, we say that $\lambda \prec_\Lambda \lambda'$ if and only if a tuple $\langle$\newepoch, \currentepoch$\rangle$ is decided for consensus instance number $n$, a tuple $\langle$\newepoch, \currentepoch'$\rangle$ is decided for consensus instance number $m$, and $n < m$.
\end{defn}

The fact that this is a total order derives directly from the following two observations.
First, all primaries are ordered: a leader becomes a primary with epoch number \currentepoch\ only after its \ready\ variable is set to true, and this occurs only after the $\langle$\newepoch, \currentepoch$\rangle$ is decided for some consensus instance.
Second, a primary $\lambda$ proposes a $\langle$\newepoch, \currentepoch$\rangle$ tuple only once and in a single consensus instance.

In the following, we use the notation $\lambda$(\currentepoch) to denote the primary with epoch number \currentepoch.
 
\begin{lemma}
\label{lemma:da-last-epoch}
If a process $p$ enters a value $v$ sent by a primary $\lambda$(\msgtentativeepoch) in $da[$\msgseqno$]$ upon deciding on instance number $n$, then  $\langle$\newepoch, \msgtentativeepoch $\rangle$ is the latest \newepoch\ tuple decided before instance $n$ and the tuple $\langle$\msgvalue, $v$, \msgtentativeepoch, \msgseqno$\rangle$ is decided for instance $n$.
\end{lemma}

\noindent {\em Proof.} Process $p$ adds $v$ to $da$ upon deciding a $\langle$\msgvalue, $v$, \msgepoch, \msgseqno$\rangle$ tuple such that \msgepoch\ $=$ \currentepoch.
The variable \currentepoch\ is set to \msgtentativeepoch\ upon deciding on the last $\langle$\newepoch, \msgtentativeepoch\ $\rangle$ tuple, so \msgepoch $=$ \currentepoch $=$ \msgtentativeepoch.$\hfill\Box$

\begin{lemma}
\label{lemma:symmetry}
If a process delivers a value $v$ upon deciding on consensus instance $n$, every process that decides on $n$ delivers $v$.
\end{lemma}

\noindent {\em Proof.} All processes decide the same values in the same consensus instance order.
A value $v$ is delivered only upon deciding on a \msgvalue\ tuple containing $v$. 
The decision of whether to deliver a value or not only depends on variables whose value is consistent across processes, since they are modified consistently by all processes every time a decision for the next instance is reached.
Therefore, every process deciding on instance $n$ delivers the same \msgvalue\ tuple and takes a consistent decision on delivering $v$.$\hfill\Box$

\begin{lemma}
\label{lemma:once}
A process delivers a value $v$ in at most one consensus instance.
\end{lemma}

\noindent {\em Proof.} Consider all values that are broadcast to be distinct.
A value $v$ can only be delivered upon deciding a \msgvalue\ tuple that contains $v$.
The primary $\lambda$ that has broadcast $v$ can include $v$ in more than one \msgvalue\ tuple.
In fact, $\lambda$ stores $v$ in $ta[li]$ every time it proposes $v$ for consensus instance $li$.
The value $ta[li]$ is proposed, however, only if $v$ is not delivered upon deciding on instance $li$.$\hfill\Box$

\begin{lemma}
Algorithm~\ref{alg:simul} satisfies Integrity: If some process delivers $v$ then some process has broadcast $v$.
\end{lemma}

\noindent {\em Proof.} A value $v$ is delivered only upon deciding on a \msgvalue\ tuple containing $v$. 
These tuples are proposed either when $v$ is broadcast, or when another \msgvalue\ tuple is decided.
In the latter case, $v$ is taken from the array $ta$.
However, only values that are broadcast are entered in $ta$.$\hfill\Box$.

\begin{lemma}
Algorithm~\ref{alg:simul} satisfies Total Order: If some process delivers $v$ before $v'$ then any process that delivers $v'$ must deliver $v$ before $v'$.\end{lemma}

\noindent {\em Proof.} Let $p_i$ deliver $v$ before $v'$ and $p_j$ deliver $v'$. 
We now show that $p_j$ delivers $v$ before $v'$. 

Processes deliver values only upon deciding on a consensus instance.
From Lemma~\ref{lemma:once}, this happens only once per value.
Since $p_i$ delivers $v$ and $v'$, this occurs upon deciding on instances $n$ and $m$ respectively, with $n < m$.
Since $p_j$ delivers $v'$, it decides on instance $m$; this happens only after deciding on instance $n$.
From Lemma~\ref{lemma:symmetry}, $p_j$ delivers $v$ upon deciding on instance $n$.$\hfill\Box$

\begin{lemma}
Algorithm~\ref{alg:simul} satisfies Agreement: If some process $p_i$ delivers $v$ and some other process $p_j$ delivers $v'$, then either $p_i$ delivers $v'$ or $p_j$ delivers $v$.
\end{lemma}

\noindent {\em Proof.} Processes deliver values only upon deciding on a consensus instance.
From Lemma~\ref{lemma:once}, this happens only once per value.
Let $n$ (resp. $m$) be the consensus instance in which $p_i$ (resp. $p_j$) decides upon and deliver $v$ (resp. $v'$).
If $n < m$, then $p_j$ decides on $n$ before deciding on $m$, so it delivers $v$ from Lemma~\ref{lemma:symmetry}.
Otherwise, $p_i$ decides on $m$ before $n$ and delivers $v'$.$\hfill\Box$

\begin{lemma}
Algorithm~\ref{alg:simul} satisfies Local Primary Order: If a primary $\lambda$ broadcasts $v$ before $v'$, then a process that delivers $v'$ delivers $v$ before $v'$.
\end{lemma}

\noindent {\em Proof.} Assume by contradiction that a primary $\lambda$ broadcasts a value $v$ followed by a value $v'$, and a process $p$ delivers $v'$ but not $v$.
Processes deliver values only upon deciding on a consensus instance.
From Lemma~\ref{lemma:once}, this happens only once per value; from Lemma~\ref{lemma:symmetry}, this happens, given a value, when deciding on the same consensus instance for all processes.
Let $n$ be the consensus instance whose decision triggers the delivery of $v'$, and let $\langle$\msgvalue, $v'$, \msgepoch, \msgseqno$\rangle$ be the tuple decided for $n$.
The value $v'$ is stored in the array $da[l]$ for some index $l$; the array $da$ was reset the last time a \newepoch\ tuple was decided, say at consensus instance $m < n$. 

Upon deciding on $m$, the value of $s$ is set to $m+1$.
After deciding on $m$ and until the decision on $n$, $s$ is incremented only when the value $da[s]$ is delivered.
Therefore, before delivering $v'$, a process has delivered all elements of $da$ with indexes in $[m+1,l]$.
From Lemma~\ref{lemma:da-last-epoch}, all values $v''$ added in $da[$\msgseqno'$]$ are sent by $\lambda$, which includes \msgseqno' in the \msgvalue\ tuple based on its counter $ls$ - this holds even if the \msgvalue\ tuple is resent more than once by $\lambda$, since the value of the \msgseqno\ field is determined by reading the original index stored in the $ta$ array.

$\lambda$ sets $ls$ to $m+1$ upon election and increments every time a new tuple is broadcast.
If $\lambda$ broadcasts \val\ before $v'$, it has sent a \msgvalue\ tuple with \msgseqno\ $\in [m+1, l-1]$. 
The value of $da[$\msgseqno$]$ held by a process after deciding on $m$ and before deciding on $l$ can only be \val.
Therefore, if a process delivers $v'$, it also delivers \val.$\hfill\Box$

\begin{lemma}
Algorithm~\ref{alg:simul} satisfies Global Primary Order: If a primary $\lambda$ broadcasts \val, $\lambda'$ broadcasts $v'$, and $\lambda \prec_\Lambda \lambda'$, then a process that delivers both \val\ and $v'$ delivers \val\ before $v'$.
\end{lemma}

\noindent {\em Proof.} By definition, if $\lambda \prec_\Lambda \lambda'$ then the \newepoch\ tuple of $\lambda$ is decided for a consensus instance $n$ preceding the one, $m$, in which the \newepoch\ tuple of $\lambda'$ is decided.
Each primary sends a \newepoch\ tuple only once.
When a process delivers $v'$, it has it stored in the $da$ array, which is reset every time a \newepoch\ tuple is delivered.
From Lemma~\ref{lemma:da-last-epoch}, $v'$ is delivered upon deciding on a consensus instance following $m$.
Similarly, a process delivers \val\ upon deciding on a consensus instance $l \in [n+1,m-1]$.
Since decisions on consensus instances are taken sequentially, \val\ is delivered before $v'$.$\hfill\Box$

\begin{lemma}
Algorithm~\ref{alg:simul} satisfies Primary Integrity: If a primary $\lambda$ broadcasts \val, $\lambda'$ broadcasts $v'$, $\lambda \prec_\Lambda \lambda'$, and some process delivers \val, then $\lambda'$ delivers \val\ before it broadcasts $v'$.
\end{lemma}

\noindent {\em Proof.} By definition, if $\lambda \prec_\Lambda \lambda'$ then the \newepoch\ tuple of $\lambda$ is decided for a consensus instance $n$ preceding the one, $m$, in which the \newepoch\ tuple of $\lambda'$ is decided.
Each primary sends a \newepoch\ tuple only once.
When a process delivers \val, it has it stored in the $da$ array, which is reset every time a \newepoch\ tuple is delivered.
From Lemma~\ref{lemma:da-last-epoch}, $v'$ is delivered upon deciding on a consensus instance $l \in [n+1, m-1]$.
The primary $\lambda'$ proposes its \newepoch\ tuple for instance $m$ after deciding on instance $l$.
By Lemma~\ref{lemma:symmetry}, $\lambda'$ delivers \val\ upon deciding on instance $l$, before sending its \newepoch\ tuple, and thus before becoming a primary and broadcasting values.$\hfill\Box$

\begin{lemma}
\label{lemma:omega-primary}
Eventually, there exists a correct primary $\lambda$ such that for any other primary $\lambda'$ ever existing, $\lambda' \prec_\Lambda \lambda$
\end{lemma}

\noindent {\em Proof.} Let $p$ be a correct process that is eventually indicated as a leader by $\Omega$. 
Upon becoming a leader, $p$ chooses a new epoch number \tentativeepoch\ and sends a \newepoch\ tuple for its current instance $d$.

Process $p$ continues sending \newepoch\ tuples for all consensus instances from $d$ on until its tuple is accepted and it becomes a primary.
We show this by induction.
For each instance $n \geq d$ in which its \newepoch\ tuple is not decided, the following can happen.
If another \newepoch\ tuple is decided with an epoch number \msgtentativeepoch\ $\neq$ \tentativeepoch, then $p$ tries to become a primary in the next instance $n+1$.
If a \msgvalue\ tuple is decided, $p$ is not yet a primary so \ready\ is false. 
Also in this case,  $p$ tries to become a primary in the next instance $n+1$.
Eventually, $p$ remains the only leader in the system.
All other processes either crash or have \ready\ set to false because they are not currently leader.
Therefore, eventually a \newepoch\ tuple from $p$ is decided, $p$ becomes the primary $\lambda$, and it is the last \newepoch\ tuple decided in any consensus instance.
According to the definition of the total order of primaries, $\lambda$ is the maximum.$\hfill\Box$.

\begin{lemma}
Algorithm~\ref{alg:simul} satisfies Eventual Single Primary: There eventually exists a correct primary $\lambda$ such that there is no subsequent primary $\lambda'$ with $\lambda \prec_{\Lambda} \lambda'$ and all messages $\lambda$ broadcasts are eventually delivered by all correct processes.
\end{lemma}

\noindent {\em Proof.} It follows from Lemma~\ref{lemma:omega-primary} that there exists a $\lambda = \lambda$(\currentepoch) that is maximum in the total order of primaries.
By definition, its $\langle$\newepoch, \currentepoch$\rangle$ tuple is the last \newepoch\ tuple decided for an instance $n$.
After deciding on $n$, $\lambda$ proposes all tuples that are broadcast using a sequence number $ls$ that starts from $n+1$ and is increased by one every time a new value is broadcast.

We now show that all tuples proposed by $\lambda$ are eventually delivered by induction on $ls$, starting from $n+1$.
Assume by induction that all values proposed by $\lambda$ with sequence number up to $ls-1$ are eventually delivered (trivially true if $ls=n+1$ since no value is proposed by $\lambda$ with a lower sequence number). 
The primary $\lambda$ proposes \val\ for $ls$ for the first time in consensus instance $li$, and sets $ta[m] = \langle v, ls \rangle$.
Note that $li \geq ls$ since $li$ is incremented every time $ls$ is.

Let $d$ be the value of $li$ at $\lambda$ when \val\ was proposed.
If $\langle$\msgvalue, \val, \currentepoch, $ls\rangle$ is decided in instance $d$, then $da[ls]$ is set to \val.
By induction, all values from $n$ to $ls-1$ are eventually included delivered, so they are added to $da$.
When all elements of $da$ with indexes in $[n+1, ls - 1]$ are filled, the element $da[ls]$ is delivered too.

If a different \msgvalue\ tuple is decided in instance $d$, then it is not from the $\lambda$, since $\lambda$ proposes only one value per instance $li$.
Therefore, \msgepoch\ $\neq$ \currentepoch, \ready\ is true for $\lambda$, and $li \geq d$, so the leader proposes a $\langle$\msgvalue, \val, \currentepoch, $ls\rangle$ tuple again, this time for a new consensus instance $li$. 
This resending can happen several time.
Eventually, however, only $\lambda$ proposes \msgvalue\ tuples, so only its tuples are decided.$\hfill\Box$

\begin{lemma}
Algorithm~\ref{alg:simul} satisfies Delivery Liveness.
\end{lemma}
\noindent {\em Proof.} Assume that a process delivers \val\ after deciding it in consensus instance $i$. By the liveness property of consensus, every correct process eventually decides \val\ in instance $i$. Since the decision on delivering \val\ deterministically depends on the consistent order of decided values, all processes deliver \val.$\hfill\Box$

The following Theorem~\ref{thm:POabcast-PC} derives directly from these lemmas and from Lemma~\ref{lem:PB}.

\begin{thm} 
\label{thm:POabcast-PC}
Barrier-free primary order atomic broadcast is correct.
\end{thm}

%
%



\end{document}